# Directional Roll-up of Nanomembranes Mediated by Wrinkling

P. Cendula,[1,2,*] S. Kiravittaya,[1] I. Mönch,[1] J. Schumann,[1] and O. G. Schmidt[1,3]

[1] *Institute for Integrative Nanosciences, IFW Dresden, Helmholtz Strasse 20, 01069 Dresden, Germany*

[2] *DCMP, Faculty of Mathematics and Physics, Charles University in Prague, 12116, Czech Republic*

[3] *Material Systems for Nanoelectronics, Chemnitz University of Technology, Reichenhainer Strasse 70, 09107 Chemnitz, Germany*

We investigate the relaxation of rectangular wrinkled thin films intrinsically containing an initial strain gradient. A preferential rolling direction, depending on wrinkle geometry and strain gradient, is theoretically predicted and experimentally verified. In contrast to typical rolled-up nanomembranes, which bend *perpendicular* to the longer edge of rectangular patterns, we find a regime where rolling *parallel* to the long edge of the wrinkled film is favorable. A non-uniform radius of the rolled-up film is well reproduced by elasticity theory and simulations of the film relaxation using a finite element method.

Bending of thin films is applied on many length scales, finding benefits both for large scale structures such as thermostats [1] and micro-/nanoscale objects including cantilevers [2], rolled-up tubes [3-4] and more recently carbon nanoscrolls formed by rolling of graphene [5,6,7]. Many promising applications of rolled-up nanomembranes have been proposed and demonstrated [8]; however each of them requires special structural and geometrical prerequisites. For instance, long tubes with small diameters are most suitable for applications such as nanopipelines [9] and metamaterial fibers [10]. Compact tubes with many windings are preferred for capacitors [11] or coils [8]. Similarly, carbon nanoscrolls possess unusual electronic [12] and optical properties [13], and because of their unique topology, they facilitate chemical doping [14] and hydrogen storage [15] for possible use in super capacitors and batteries [14]. Rigorous controllability of the rolling process and the realization of novel tubular shapes lie at the heart of a more deterministic deployment of these rolled-up nanomembranes.

In this Letter, we discuss a new approach to control the rolling process of nanomembranes by introducing wrinkles to the initial film. These nanomembranes are not limited to typical thin films but they also include two-dimensional atomic crystals such as graphene. Analogous to the macroscopic scale, where wrinkles are commonly taken to modify the elastic properties of a layered structure (e.g. cardboard paper), we can apply wrinkles to a thin solid film in order to exert control over the rolling direction once the film is released from a rectangular substrate pattern. We theoretically investigate the relaxation of strained rectangular wrinkled films, taking into account variations of strain gradient, wrinkle geometries and pattern sizes. Our theory predicts a preferential rolling direction based on elastic energy and anticipates a non-uniform radius of rolled-up tubes obtained from the initial wrinkled film. Finally, experimental data are provided which confirm our theoretical predictions and open up new possibilities to realize and control 3D tubular structures on many length scales with well-defined geometry and broad application potential.

Let us consider a typical flat rectangular film with initial strain gradient $\Delta\varepsilon$ across the film thickness. The strain gradient might originate from i.e. mismatch strain [3], thermal strain [16] or surface stress imbalance in ultrathin layers of graphene [6] or silicon [17]. Once the film is released from the substrate, it first bends at all four edges simultaneously. However, further bending (rolling) occurs only along the long edges of the rectangular film, because the longer edges experience larger bending force/moment than the shorter edges [18]. To experimentally overcome the limitation of tubes forming only at the long edge of the pattern together with poor rolling behavior at the corners and to control the final position of the tubes, shallow etch [19] and angle deposition techniques [16] have been developed. However, introducing wrinkles to the initial film is an alternative to overcome drawbacks of those methods, which include film rupture at pattern edges, damage by inaccurate etching and traces of auxiliary layers.

When a strained wrinkled film patterned into rectangular shape, as shown in Fig. 1 (a), is gradually released from its substrate (either by means of chemical reaction or by removing an underlying sacrificial layer) elastic relaxation occurs by bending and rolling-up the film either into a tube at the wrinkled edge (TWE) [Fig. 1(b)] or into a tube at the flat edge (TFE) [Fig. 1(c)]. We show that bending the wrinkled edge (WE) can be efficiently suppressed by an energy barrier arising from the need to flatten the released wrinkled film during bending. The relaxation of the released part of the film, as denoted in Figs. 1 (b) and (c), is described by the change in total elastic energy between the initial film and the relaxed WE or flat edge (FE). For this purpose, we compute the change in elastic energy of the relaxed film portion on the WE and FE, followed by the computation of the total energy change of the released film area from these two contributions.

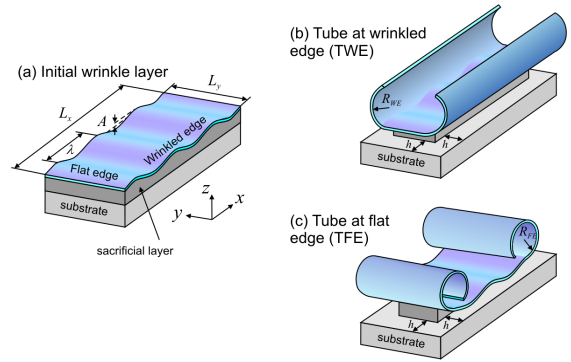

FIG. 1 Sketch of (a) initial rectangular wrinkled film. The coordinate system and notation of flat and wrinkle edges are defined. Upon release of the film over a length *h*, the film relaxes by either (b) rolling-up at the wrinkled edge with a curvature radius $R_{WE}$ or (c) rolling-up at the flat edge with the curvature radius $R_{FE}$.

The initially rectangular wrinkled film of size $L_x$ and $L_y$ and thickness $t$ is approximated by a sinusoidal height profile $\zeta_0 = A\sin(kx)$ [see Fig. 1(a)], where $A$, and $k$ are the amplitude and wavenumber of the wrinkle in the $x$-direction. We consider small slope wrinkles, where the amplitude $A$ is much smaller than the wrinkle wavelength $\lambda$. It is most interesting to consider a film which is longer in the wrinkle direction, i.e., $x$-direction ($L_x > L_y$) since wrinkling along the shorter direction only enhances the tendency of the film to roll-up into that very same direction. The initial curvature of the wrinkled film is $c_0(x) \approx -Ak^2\sin(kx)$, and the initial biaxial in-plane strain is $\varepsilon_0 = \bar{\varepsilon} + (\Delta\varepsilon/t)(z-\zeta_0)$, (which is linear through the film thickness), where $\bar{\varepsilon}$ and $\Delta\varepsilon$ are average strain and strain gradient of the film, respectively. Isotropic linear elastic properties of the film with Young's modulus $E$ and Poisson's ratio $\nu$ are applied throughout.

Etching the sacrificial layer leaves the film of length $h$ on all four edges free to relax its energy. The bending moment per unit width of the film portion $M_0 = (1+\nu)D\Delta\varepsilon/t$ (equal for WE and FE) can be derived by integrating the normal force through the thickness of the film, where $D = Et^3/[12(1-\nu^2)]$ is the bending rigidity of the flat film. The bending (roll-up) radius of the flat film portion $R_0$ due to the bending moment $M_0$ can be calculated by the classical expression for pure bending of a thin plate [20] $R_0 = D/M_0 = t/((1+\nu)\Delta\varepsilon)$ and the corresponding curvature is $g_0 = 1/R_0$.

We evaluate the change in elastic energy of the released film portion undergoing attenuation of wrinkles and transition to the rolled-up state (in $y$-direction) by a simple bi-stability model [21]. The curvature of the film is separated into a 'local' wrinkle curvature $c(x)$ and a 'global' curvature $g$ of the film. We assume that the wrinkle profile remains sinusoidal during rolling. The height profile along the curved surface is defined as $\zeta(\gamma) = (1-\gamma)\zeta_0$. This profile can be attenuated with an attenuation parameter $\gamma$ while the wrinkle period is kept constant. The value of the attenuation parameter $\gamma = 0$ corresponds to the initial wrinkle amplitude and $\gamma = 1$ denotes a film that is locally flat.

The initial total curvature can be written in a vector notation [21], i.e., $\kappa_0 = (c_0(x),0)$ and the actual curvature during the wrinkle attenuation process and rolling up is $\kappa = (c(x),g)$, where $c(x) = (1-\gamma)c_0(x)$. Note that even for $\gamma = 0$ there is some global curvature $g$. The change in the total curvature between these two states is $\Delta\kappa = \kappa - \kappa_0 = (-\gamma c_0(x), g)$. The average change in elastic energy density of the wrinkled film per period caused by the curvature change $\Delta\kappa$ is given by

$$\Delta\bar{u}_{WE} = -M_0 g + \frac{1}{2}D\left((1+\alpha_\gamma)g^2 + \gamma^2(Ak^2)^2/2\right), \quad (1)$$

where $\alpha_\gamma = 6(1-\nu^2)(1-\gamma^2)A^2/t^2$ includes the contribution from the different bending rigidity along the wrinkles [22]. Equilibrium with respect to $g$ is obtained as $g_{Eq}(\gamma) = M_0/[D(1+\alpha_\gamma)]$. If the wrinkles are almost fully attenuated ($\gamma = 1$), the radius of the tube at the WE is given by the reciprocal value of the equilibrium curvature $R_{WE} = 1/g_{Eq}(1) = R_0$. Inserting this equilibrium curvature into Eq. (1), we obtain

$$\Delta u_1(\gamma) \equiv \Delta\bar{u}_{WE,Eq} = -\frac{M_0^2}{2D(1+\alpha_\gamma)} + \frac{1}{4}D\gamma^2(Ak^2)^2. \quad (2)$$

To simplify the notation, we introduce $\Delta u_1(\gamma)$, which is a function of $\gamma$ and always negative implying that the rolling at the WE is an energetically favorable process. We numerically search for the maxima of energy change $\Delta u_1(\gamma)$ and identify three characteristic energy regimes as shown in the insets of Fig. 2. These regimes are used when considering the total energy change after analyzing rolling at the FE.

Rolling of the film portion at the FE is now considered. The curvature in $x$ direction is assumed to change abruptly from the initial curvature $c_0(x)$ to the curvature of the rolled-up tube $g_{FE}$ (reciprocal of radius: $R_{FE}$), therefore $\Delta\kappa = (g_{FE} - c_0(x), 0)$. The resulting change in energy is calculated similarly as for the WE and we can easily obtain

$$\Delta u_{FE} = -M_0\left(g_{FE} + k^2\zeta_0 - R_0(g_{FE} + k^2\zeta_0)^2/2\right). \quad (3)$$

The equilibrium with respect to global curvature $g_{FE}$ leads to the dependence of the tube curvature radius on the reference coordinate $x$

$$g_{FE}(x)/g_0 = 1 - R_0 k^2 A\sin(kx), \quad (4)$$

with the corresponding energy $\Delta u_{FE,Eq} = -M_0^2/(2D) \equiv \Delta u_2$, where we introduce a simplified notation $\Delta u_2$. Since $\Delta u_2 < 0$, rolling-up TFE is always energetically favorable as compared to the unrolled film.

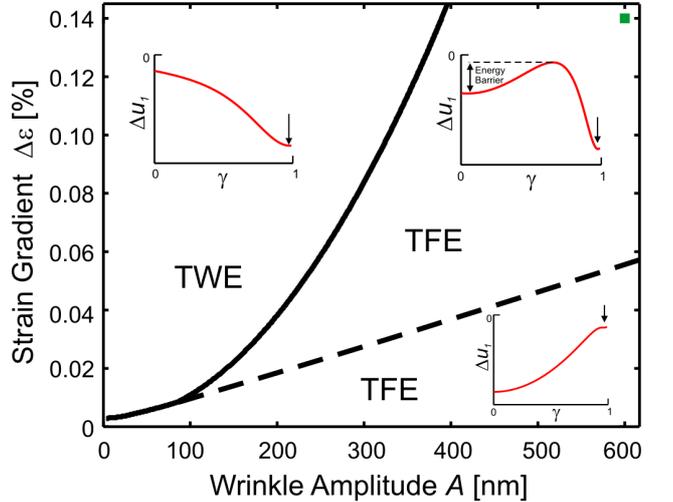

FIG. 2. Phase diagram of tube shapes after release of a rectangular patterned wrinkled film as a function of wrinkle amplitude $A$ and initial strain gradient $\Delta\varepsilon$. Three regions are identified based on the characteristics of the calculated elastic energy landscape. On the left side the film rolls up at the wrinkled edge (TWE), whereas for the two regions on the right side the film starts curling into tubes from the flat edge (TFE). Insets show the elastic energy change as a function of wrinkle attenuation parameter $\gamma$. Elastic properties of CuNi alloy (Constantan) $E = 162$ GPa, $\nu = 0.33$ [23] is assumed for this calculation. A film thickness $t = 70$ nm and a wrinkle wavelength $\lambda = 40$ μm are considered. These parameters were chosen in view of our experimental result, which is shown as a green data point in the upper right corner.

Up to here, we considered a portion of the released film bending in either of the 2 directions (WE, FE). We can now consider the change in total energy of the whole released film (consisting of four edges) and the competition between TWE and TFE. Note that we do not consider relaxation at the corners in this simplification.

In the regime, left of the solid curve in Fig. 2, the energy change $\Delta u_1$ typically decreases with increasing $\gamma$ and the minimum point is near $\gamma = 1$ (arrow in the inset). If TWE is to prevail, we can approximate that the film on the FE is unrelaxed with $\Delta u = 0$, see Fig. 1(b), and therefore the total energy change originates only from the rolling at the WE, $\Delta U_1 = 2\Delta u_1(1) L_x h$, where $h$ is the release length. If TFE should prevail, we approximate that the film at the WE is unrelaxed and the relaxation comes only from the rolling at the FE, $\Delta U_2 = 2\Delta u_2 L_y h$. The rolling is preferred energetically at the WE when it releases more energy, $\Delta U_1 < \Delta U_2$, and thus we obtain a condition for the aspect ratio of the stripe $L_y/L_x < 1 - (R_0 A k^2)^2/2$, for which the tube rolls up from the WE. This ratio is always smaller than 1 and equals zero for $A_{max} = \sqrt{2}/(R_0 k^2)$. Therefore, beyond this maximum wrinkle amplitude, one cannot find any aspect ratio $L_y/L_x$ to make TWE energetically preferable.

In the regime below the dashed curve in Fig. 2, forming TWE would require an increase of energy to $\Delta U_1$ (inset), which is not favorable. In contrast, TFE can be formed since the energy $\Delta U_2 < 0$ is released. The TWE is equal in energy to the initial wrinkled film (dashed curve) for $\Delta u_1(0) = \Delta u_1(1)$, resulting in the scaling $At/\lambda^2$ ($A \gg t$) for the phase boundary of these regions.

Between the solid and dashed curves, the film at the WE has to increase its energy from $\Delta u_1(0)$ to the maximum of $\Delta u_1(\gamma)$ (and also the total energy) to progress into the rolled-up state near $\gamma = 1$ (arrow) (see upper right inset of Fig. 2). Under equilibrium conditions, the film cannot overcome this energy barrier and change its shape without some external forces. Therefore TWE is suppressed and TFE is formed, because $\Delta U_2 < 0$. It should be noted that even though the final shape is the same (TFE) as below the dashed curve, the film might be snapped to the second stable TWE state by external forces, which can be generated for instance during the fabrication process (e.g., capillary forces, flowing pressure). The solid curve in Fig. 2 separates the region without energy barrier on the left from that with an energy barrier on the right and it scales as $A^2/\lambda^2$ [24]. Interestingly, this scaling is thickness independent and varies only with wrinkle parameters $A$, $\lambda$. For a given strain gradient, i.e. $\Delta \varepsilon = 0.1\%$ in Fig. 2, there exists a minimum wrinkle amplitude $A_m = 330$ nm for suppression of the rolling at the WE (solid curve). We consider scrolling of a wrinkled graphene sheet within our continuum elasticity approach. Based on recent experiments [6,7], the approximate radius of scrolled graphene is $R_0 = 30$ nm and the effective graphene thickness is $t = 0.34$ nm, which leads to an effective strain gradient $\Delta \varepsilon$ of 1.1%. By a similar calculation which produced Fig. 2, we predict that wrinkling with an amplitude of 3 nm can suppress scrolling from the wrinkled graphene edge. Such predictions are useful in the design of wrinkle parameters for directional rolling and might help to control the chirality of rolled-up nanotubes.

To confirm that our analytical Eq. (4) is suitable to describe the curvature of the TFE, we used numerical finite element method with large deformations [26] to compute the relaxation of the strained wrinkled film. For simplicity, we used a 2D cross-section of the film with five periods and linearly increased the initial strain in the whole film rather than dynamically removing the sacrificial layer and releasing the film gradually [27]. The wrinkled film is fixed at the etching front and free otherwise and its relaxed non-uniform tube shape is shown in Fig. 3(a). The normalized curvature $g_{FE}(x)/g_0$ is extracted from the relaxed shape in Fig. 3(a) and it compares very well to Eq. (4) in Fig. 3(b). Even though the normalized curvature oscillates around the curvature of the flat film $g_0$ and can be even negative in some intervals, the overall shape of the relaxed film is still tubular with an average curvature $R_0$. By proper design of the wrinkle profile, which matches the circumference of the tube to the integer multiple of wavelength, one can obtain new tubular structures with varying curvature along the azimuthal direction.

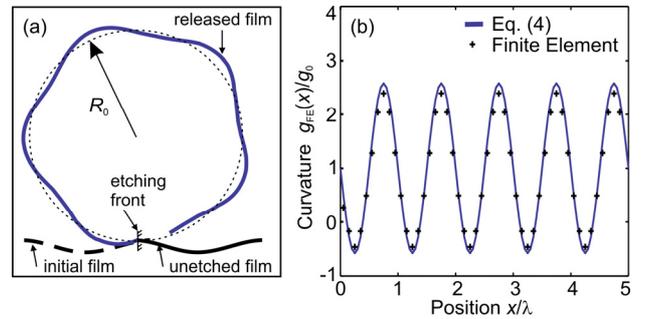

FIG. 3. Finite element study of rolling the wrinkled film from the flat edge with 5 wrinkle periods, $\Delta \varepsilon = 2.4\%$, $A = 100$ nm and $\lambda = 2$ μm. (a) Almost complete rotation of the released film with non-uniform tubular shape (solid blue line) and uniform tube dashed line) as a guide to the eyes. (b) Curvature of the tube in (a), where the solid line is generated by the theoretical calculation using Eq. (4) and crosses show the results extracted from finite element calculations.

In order to confirm our theoretical predictions, strained wrinkled films were fabricated by deposition of metallic CuNiMn alloyed material on wrinkled surfaces of photoresist. There are plenty of techniques to fabricate wrinkled patterns such as buckling of patterned films on elastomer substrate [29], uniaxial stretching of sheets [25,29], gray-scale lithography [30] or chemical patterning of substrates with regions of high and low adhesion [31]. Here, we have prepared wrinkled photoresist surfaces by multi-exposure steps. Two kinds of wrinkle profiles, sinusoidal 'sin' and 'step' with a wavelength of 40 μm, together with a flat profile for comparison were fabricated. The metallic $Cu_{0.59}Ni_{0.40}Mn_{0.01}$ alloy films with thicknesses of 70 or 100 nm (±5 nm) were deposited on top of the wrinkled photoresist by magnetron sputtering at room temperature. The morphology of the wrinkle profiles are characterised by atomic force microscopy (AFM) as shown by their cross-section profile in Fig. 4(a). The measured 'sin' profile is well fitted by a sinusoidal function with an amplitude of 600 nm (tuned by exposure time). A second lithographic step was performed to fabricate rectangular 180×360 μm² patterns.

The samples with rectangular flat and wrinkled films are covered by droplets of N-methyl-2-pyrrolidone in order to gradually remove the sacrificial photoresist layer. Immediately after putting the sample into the liquid, we start recording the film releasing process. Images from etching the flat and wrinkled films are shown for comparison in Figs. 4(b) - (e). The flat film initially bends at all four edges [Fig. 4(b)], and as etching proceeds, roll-up of the longer edge prevails and tubes with their axis along the long edge form as we theoretically described above. The strain gradient in the deposited film was estimated from the average radius of the tubes from the flat film measured in the etchant liquid after total release, $R_0 = 36 \pm 5\ \mu m$, which directly yields $\Delta\varepsilon = -0.14 \pm 0.02\%$. In contrast, the wrinkled film with the 'sin' profile starts to bend preferentially at the flat edges without any noticeable bending at the wrinkled edges [see Fig. 4(c)], and finally the tube is formed along the flat edge. All 100 rectangular patterns monitored reveal the same behaviour. Taking the computed strain gradient and wrinkle amplitude of 600 nm, we see that the observed rolling direction is in agreement with our phase diagram in Fig. 2 (marked by the square). The tubes which rolled from the wrinkled film apparently experience a larger radius (51±5 μm) than the tubes rolled from the flat film (36±5 μm). We explain this obervation by the inhomogeneous tube radius as predicted in Eq. (4). The 'step' wrinkle profile behaves in a similar manner during release, forming the tubes along the flat edge [Fig. 4(e)]. Therefore, the functional form of the wrinkle profile can be varied without any significant effect on the control of the rolling direction. Our experimental method might be used for the deterministic rolling of ultimately thin films such as graphene into well-defined chiralities with applications in electronics as well as actuation and lab-on-chip devices [33].

In summary, we have theoretically investigated the roll-up of nanomembranes formed by releasing wrinkled rectangular films. The introduction of wrinkles can be used to control the preferential rolling direction of strained films as was confirmed experimentally by depositing and releasing strained metallic alloys on sinusoidal/step-like photoresist patterns. Given the abundance of fabrication methods and applications of wrinkled [29] and rolled up [34] films across length scales (from conventional thin films down to graphene), our work will be useful for realizing novel 3D tubular structures with well-controlled geometry.

The authors acknowledge B. Eichler and U. Nitzsche for technical assistance and A. D. Norman, E. J. Smith, J. D. A. Espinoza, D. Grimm, C. Deneke and Y. F. Mei for fruitful discussions. This work was supported by the Volkswagen Foundation (I/84 072) and the U.S. Air Force Office of Scientific Research MURI program under Grant FA9550-09-1-0550.

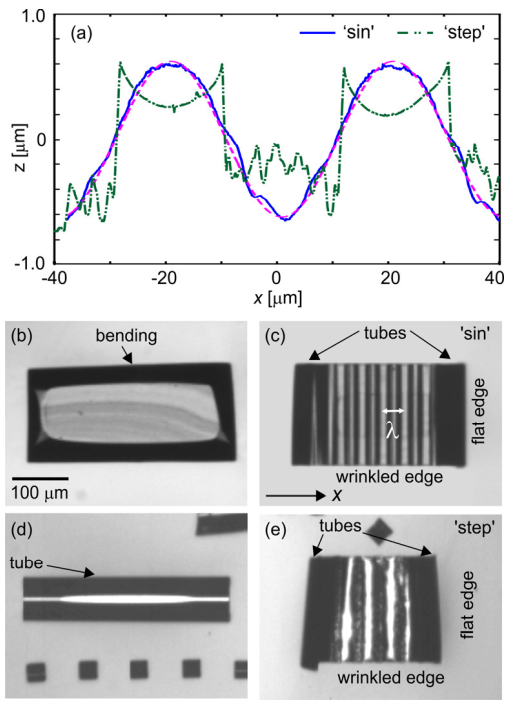

FIG. 4. (a) Cross-sectional AFM profiles of the 'sin' pattern (solid blue line) and of the 'step' pattern (dash-dotted green line). The sinusoidal profile obtained from the best fit is also shown as a dashed cyan line. (b)-(e) Snapshots from video microscopy recorded during release of the metallic film of thickness $t$ = 70 nm after ~ 1 minute under-etching of (b) a flat pattern, (c) a wrinkled 'sin' pattern, (d) another flat film with $t$ = 100 nm and (e) a wrinkled 'step' pattern. Movie versions of (b) and (c) are available [32].